\documentclass[sigconf,natbib=true]{acmart}
\AtBeginDocument{%
  }

\usepackage{bm}
\usepackage{array}
\usepackage{multirow}
\usepackage{subcaption}
\usepackage{graphicx}  % 提供 \rotatebox
\usepackage{tabularx}
\usepackage{algpseudocode}
\usepackage[ruled,vlined]{algorithm2e}
\usepackage{enumitem}
\usepackage{soul}

\newcommand{\model}{TIDE}% package

\copyrightyear{2026}
\acmYear{2026}
\setcopyright{cc}
\setcctype{by}
\acmConference[SIGIR '26]{Proceedings of the 49th International ACM SIGIR Conference on Research and Development in Information Retrieval}{July 20--24, 2026}{Melbourne, VIC, Australia}
\acmBooktitle{Proceedings of the 49th International ACM SIGIR Conference on Research and Development in Information Retrieval (SIGIR '26), July 20--24, 2026, Melbourne, VIC, Australia}
\acmDOI{10.1145/3805712.3809615}
\acmISBN{979-8-4007-2599-9/2026/07}

\begin{document}

\title{Time-Interval-Aware Disentangled Expert Modeling for Next-Basket Recommendation}

\author{Zhiying Deng}
\orcid{0000-0002-7554-0920}
\affiliation{
	\institution{Laboratory for Artificial Intelligence and New Forms of Education, Central China Normal University}
	 \city{Wuhan}
	\country{China}}
\email{zhiyingdzy@ccnu.edu.cn}
%%%
\author{Yuan Fu}
\orcid{0009-0000-7087-4975}
\affiliation{
	\institution{School of Software Engineering, Huazhong University of Science and Technology}
	\city{Wuhan}
	\country{China}}
\email{fuyuan@hust.edu.cn}
%%%
\author{Usman Farooq}
\orcid{0009-0002-3776-2685}
\affiliation{
	\institution{School of Computer Science and Technology, Huazhong University of Science and Technology}
	\city{Wuhan}
	\country{China}}
\email{usman@hust.edu.cn}
%%%
\author{Ziwei Tian}
\orcid{0009-0008-8626-3798}
\affiliation{
	\institution{School of Computer Science and Technology, Huazhong University of Science and Technology}
	\city{Wuhan}
	\country{China}}
\email{stephentian@hust.edu.cn}
%%%
\author{Wei Liu}
\orcid{0000-0002-3871-9454}
\authornote{Corresponding author.}
\affiliation{
	\institution{School of Computer Science and Technology, Huazhong University of Science and Technology}
	\city{Wuhan}
	\country{China}}
\email{idc\_lw@hust.edu.cn}
%%%
\author{Jianjun Li}
\orcid{0000-0002-5265-7624}
\affiliation{
	\institution{School of Computer Science and Technology, Huazhong University of Science and Technology}
	\city{Wuhan}
	\country{China}}
\email{jianjunli@hust.edu.cn}

\renewcommand{\shortauthors}{Zhiying Deng et al.}

\begin{abstract}
	
Next-basket recommendation (NBR) is a type of recommendation that aims to predict a set of items a user will purchase based on their historical transaction basket sequences. It is governed by a dynamic interplay between two distinct user intents: habitual repurchase, which involves repeating past behaviors, and exploratory interest, which involves discovering new items. However, existing NBR methods generally suffer from two limitations: (1) they often entangle these conflicting motives within a single representation, causing habits to overshadow discovery, and (2) they rely on discrete sequential modeling that ignores continuous-time intervals and item-specific periodicities. In this paper, we propose a novel solution named \textbf{T}ime-\textbf{I}nterval \textbf{D}isentangled \textbf{E}xperts ({\textbf{\model}}) to address these challenges. {\model} incorporates a Hawkes-enhanced Fourier Time Encoding to capture item-specific temporal periodicities and dynamic decay. To decouple user intentions, {\model} utilizes a dual-expert architecture that integrates a Habit Expert for recurring needs and a Pattern-Guided Exploration Expert for discovery. Combined with an item-aware gating mechanism, {\model} adaptively balances repurchase and exploration. Extensive experiments on four diverse real-world datasets demonstrate that {\model} consistently outperforms representative state-of-the-art NBR methods.
	
\end{abstract}

\begin{CCSXML}
	<ccs2012>
	<concept>
	<concept_id>10002951.10003317.10003347.10003350</concept_id>
	<concept_desc>Information systems~Recommender systems</concept_desc>
	<concept_significance>500</concept_significance>
	</concept>
	<concept>
	<concept_id>10010405.10003550.10003555</concept_id>
	<concept_desc>Applied computing~Online shopping</concept_desc>
	<concept_significance>500</concept_significance>
	</concept>
	</ccs2012>
\end{CCSXML}

\ccsdesc[500]{Information systems~Recommender systems}
\ccsdesc[500]{Applied computing~Online shopping}

\keywords{Next-Basket Recommendation, Habit and Exploration, Temporal Dynamics}

\maketitle

\section{Introduction}

\begin{figure}[t]
	\centering
	\includegraphics[width=\linewidth]{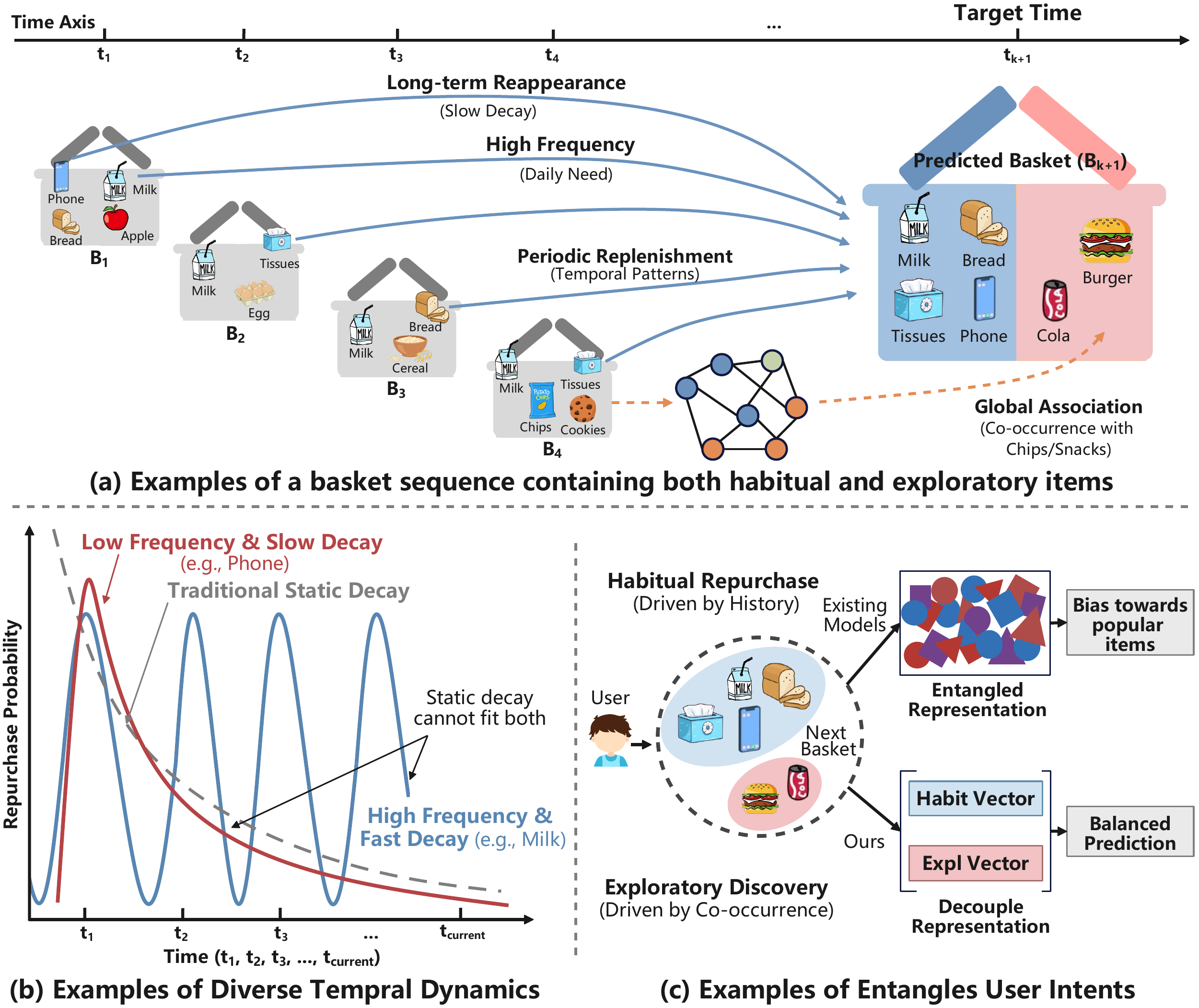}
	\caption{Motivation of {\model}. (a) The NBR task aims to predict the next basket at the target time $t_{k+1}$ based on a historical basket sequence; (b) Diverse temporal dynamics demonstrating the necessity of capturing item-specific periodicities; (c) Entangled user intents revealing that conflicting consumption motives require disentangled representation learning.}
	\label{fig_intro}
\end{figure}

Next-Basket Recommendation (NBR) has become a cornerstone of modern e-commerce and online retail systems~\cite{FPMC_2010,ReCANet_2022,mao2024ranking,liu2023retrieval,liu2025queries,shi2025answering}. As illustrated in Figure~\ref{fig_intro}(a), the primary objective of NBR is to predict a set of items (a \textit{basket}) that a user is likely to purchase at a specific target time $t_{k+1}$, given their historical transaction sequence. Unlike traditional sequential recommendations that focus on item-to-item transitions, NBR is significantly more complex due to the intricate dependencies within and between baskets. A user's trajectory in NBR often encompasses long-term item reappearance, high-frequency consumption, and periodic replenishment cycles. These characteristics distinguish NBR from conventional sequence modeling, as the \textit{basket} structure necessitates a deeper understanding of recurring purchase behaviors rather than just one-off exploratory transitions.

To capture the sequential relationships between these baskets, a common approach is the use of static decay mechanisms~\cite{TIFUKNN_2020,PengM22023,deng2023multi}. These methods assume that a user’s interest in an item is at its peak immediately after a purchase and naturally fades over time, thus assigning higher weights to the most recent interactions. While simple and widely adopted, this monotonic assumption is overly naive. With the rise of deep learning, Recurrent Neural Networks (RNNs) and Transformers have been introduced to NBR~\cite{DREAM_2016,HuSets2Sets2019,yu2023basket} to model complex dependencies. These sequence-based models (such as RNNs) and static decay methods (gray dashed line) typically follow the assumption of monotonic decay, assuming that the contribution of earlier interactions to the current ones decreases exponentially. However, they fail to perceive the non-monotonic temporal dynamics of different items, limiting their ability to predict exactly \textit{when} a user will replenish a specific item. As shown in Figure~\ref{fig_intro}(b), a durable good (e.g., a phone, the red line) may have a single long-term peak, while a daily necessity (e.g., milk, the blue line) exhibits strong, non-monotonic periodicity.

Furthermore, a unique and critical aspect of the NBR task is the presence of repeat purchase behavior. Previous studies often treat these repetitions as a weight or bias, assuming that frequently interacted items represent the core of user interest~\cite{ReCANet_2022,deng2023multi,PengM22023}. However, this has been proven to make recommendations heavily biased toward popular or repetitive items~\cite{liu2025repeat,DengLLZ25}, overshadowing the user's latent need for discovery. Recently, some works have begun to investigate the balance between repetition and exploration~\cite{li2024we,liu2025repeat}; yet, they typically entangle these conflicting motives within a single representation tower. As illustrated in Figure~\ref{fig_intro}(c), such entanglement forces habitual signals (Habit) and novel discoveries (Exploration) to compete for the same latent space. 
This lack of disentanglement leads to suboptimal performance, 
preventing the model from accurately capturing the user's shifting intent or potential interest in new categories.

To address these challenges, we propose a novel time-interval disentangled experts framework for next-basket recommendation, abbreviated as TIDE. {First}, to accurately model diverse temporal rhythms, we devise a Hawkes-enhanced Temporal Modeling module with Fourier Time Encoding. This approach goes beyond monotonic decay by jointly characterizing personalized periodicities and dynamic intensity jumps, allowing the model to learn habit repurchase. {Second}, we design a Dual-expert Architecture to explicitly decouple habitual and exploratory motives. We employ a Habit Expert to capture historical replenishment patterns and a Pattern-Guided Exploration Expert to mine potential interests from a global item-pattern bipartite graph. An item-aware gating mechanism then dynamically fuses these two views to provide a balanced and personalized recommendation.
The main contributions of this work are summarized as follows:

\begin{itemize} 
	\item We identify the limitation of conventional monotonic temporal decay and uncover the critical role of item-specific, non-monotonic periodicities in NBR, providing new insights into modeling real-world replenishment cycles.
	\item We propose the TIDE model, which incorporates a Hawkes-enhanced Fourier Time Encoding to jointly capture personalized temporal periodicities and utilizes a dual-expert architecture to decouple user intentions into habitual and exploratory components for fine-grained interest learning, which demonstrates that decoupling these distinct motives is beneficial for user interests.
	\item Extensive experiments have been conducted on four widely used benchmark datasets to evaluate the effectiveness of our model. The results show that {\model} significantly outperforms several representative state-of-the-art NBR baselines in diverse scenarios.
\end{itemize}

\section{Related Work}
\label{sec:related}
We review three lines of research that are most relevant to our work: (1) Next-Basket Recommendation, (2) Temporal Modeling in Recommendation, and (3) Disentangled Representation Learning.

\noindent\textbf{Next-Basket Recommendation.}
Early works~\cite{FPMC_2010,NN_Rec_2015,HRM_2015} utilize Markov Chains to capture local transitions between adjacent baskets, while modern NBR models leverage RNNs~\cite{DREAM_2016,CLEA_2021,yu2023basket}, attention mechanisms~\cite{Intention2Basket_2020,IntentionNets_2020}, and auxiliary signals—such as hypergraphs~\cite{zhou2025dual}, contrastive learning~\cite{wei2024knowledge}, and counterfactual inference~\cite{li2024corerec}—to model long-range dependencies~\cite{shi2025personax}. A defining characteristic of NBR is the prevalence of repeated behaviors, which many studies prioritize via nearest-neighbor modeling~\cite{TIFUKNN_2020} or pattern extraction~\cite{ReCANet_2022} to boost accuracy.
However, there is a growing consensus that habitual repetition and exploratory discovery represent fundamentally distinct motives. Most current methods~\cite{deng2023multi,li2024we} entangle these conflicting intentions within a single representation space, often allowing dominant habit signals to overshadow subtle exploratory interests. Unlike studies that rely on post-processing to balance these signals~\cite{liu2025repeat}, {\model} introduces a gated dual-expert architecture to explicitly decouple user intents at the architectural level, ensuring that habit and exploration signals are learned in separate expert modules rather than competing within a shared representation.

% Table starts
\setlength{\tabcolsep}{6pt}
\begin{table*}[t]
	\centering
	\caption{Statistics of the four real-world datasets after pre-processing.}
	\vspace{-0.5em}
	\label{tab:dataINFO}
	\resizebox{\linewidth}{!}{
		\begin{tabular}{lrrrrrrr}
			\hline
			{Dataset} & {\#Interactions} & {\#Users} & {\#Baskets} & {\#Items} & {Avg. Baskets per User} & {Avg. Items per Basket} & {Avg. Items per User} \\
			\hline
			Beauty  & 54,692  & 2,134  & 26,617  & 1,694 & 12.47 & 2.05 & 25.63 \\
			Sports  & 80,954  & 4,094  & 29,589  & 4,726 & 7.23  & 2.74 & 19.77 \\
			Grocery & 136,840 & 4,349  & 51,839  & 2,487 & 11.92 & 2.64 & 31.46 \\
			Home    & 256,221 & 11,886 & 101,066 & 9,473 & 8.50  & 2.54 & 21.56 \\
			\hline
		\end{tabular}
	}
\end{table*}

% Figure starts
\begin{figure*}[t]
	\centering
	% (a) pic1
	\begin{subfigure}[b]{0.24\linewidth}
		\centering
		\includegraphics[width=\linewidth]{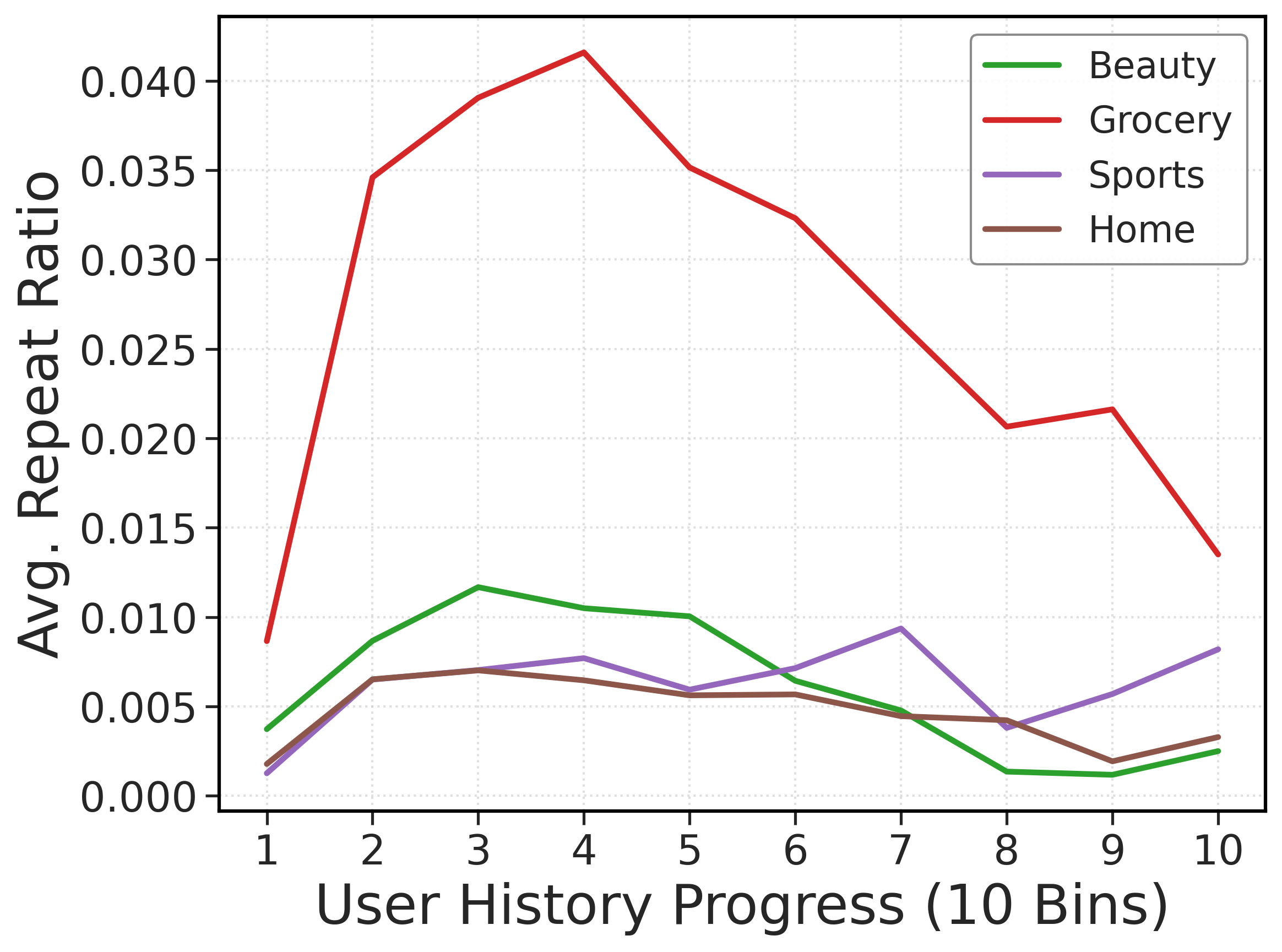}
		\caption{Repeat Ratio} 
		\label{fig_pic1}
	\end{subfigure}
	\hfill
	% (b) pic2
	\begin{subfigure}[b]{0.24\linewidth}
		\centering
		\includegraphics[width=\linewidth]{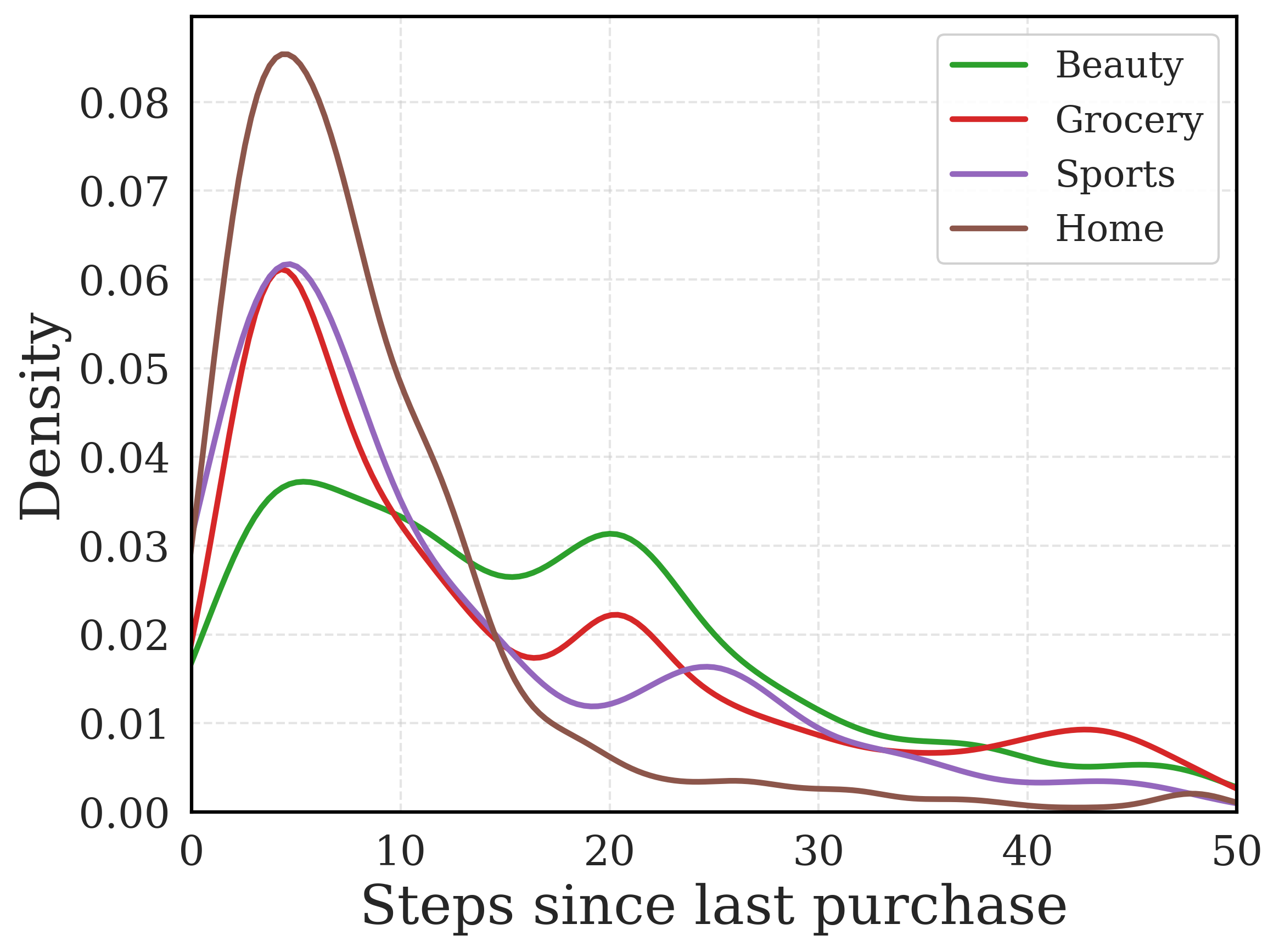}
		\caption{Interval Density} 
		\label{fig_pic2}
	\end{subfigure}
	\hfill
	% (c) pic3
	\begin{subfigure}[b]{0.24\linewidth}
		\centering
		\includegraphics[width=\linewidth]{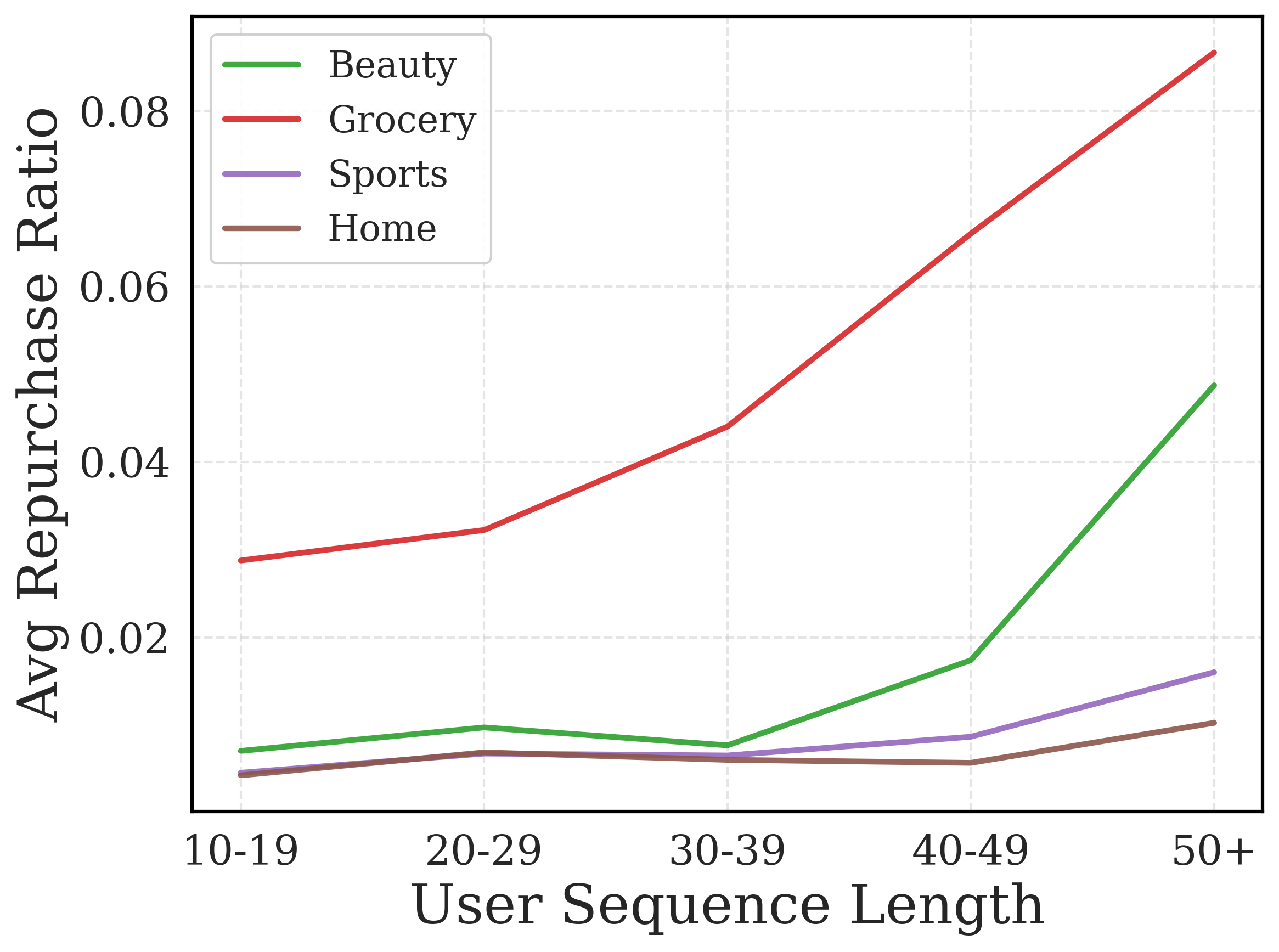}
		\caption{History Impact} 
		\label{fig_pic3}
	\end{subfigure}
	\hfill
	% (d) pic4
	\begin{subfigure}[b]{0.24\linewidth}
		\centering
		\includegraphics[width=\linewidth]{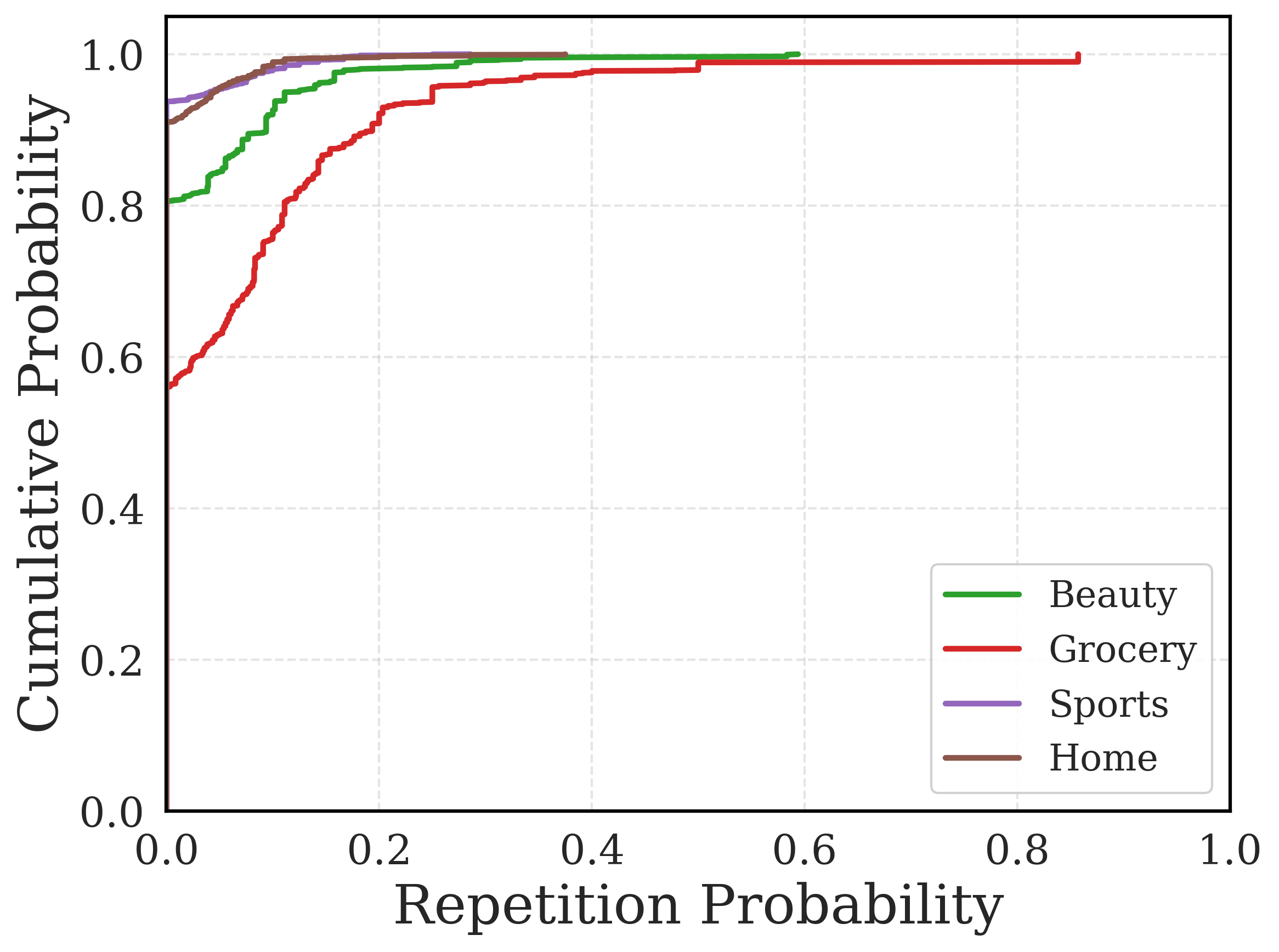}
		\caption{Item Probability} 
		\label{fig_pic6}
	\end{subfigure}
	\vspace{-0.5em}
	\caption{Statistical analysis of user consumption behaviors. (a) Evolution of Repeat Ratio. (b) Distribution of Repurchase Intervals. (c) Impact of History Length on Repurchase. (d) Item-Level Repetition Probability.}
	\label{fig_data}
\end{figure*}

\noindent\textbf{Temporal Modeling in Recommendation.}
Modeling temporal dynamics is fundamental to capturing the evolution of user preferences in sequential recommendation. Beyond simple chronological order, recent works integrate explicit temporal information via multi-dimensional embeddings \cite{rahmani2023incorporating}, time-interval augmentation \cite{dang2023uniform}, and interval-aware weighting in linear models \cite{park2025temporal}. Advanced architectures, such as Temporal Graph Transformers \cite{xia2022multi} and temporal-aware GNNs \cite{zhang2024temporal, peng2025tagrec}, capture complex transition patterns and multi-behavior correlations through contrastive learning.
In addition, frequency-domain methods have also emerged as a promising direction. \cite{shin2024attentive} introduce Fourier-based inductive biases into the self-attention mechanism to capture global periodicity in item sequences, while \cite{baek2025muffin} propose user-adaptive frequency filters to model personalized frequency components in sequential recommendation.
Despite these advances, existing methods struggle with NBR because they often assume homogeneous monotonic decay \cite{ou2025ls, dang2023uniform}, failing to capture item-specific periodicities (e.g., the short-term replenishment of \textit{milk} vs. the long-term decay of a \textit{phone}). Frequency-domain methods~\cite{shin2024attentive,baek2025muffin} apply spectral analysis over item sequences in Sequential Recommendation (SR), yet they do not account for the non-monotonic intensity dynamics unique to NBR, where a repurchase event triggers an excitation spike rather than a smooth sinusoidal oscillation. In contrast, {\model} combines Fourier Time Encoding with a Hawkes-inspired item-specific decay kernel to jointly model personalized periodicities and dynamic intensity jumps, directly targeting the \textit{when to replenish} challenge inherent to basket-level purchasing behavior.

\noindent\textbf{Disentangled Representation Learning.}
Disentangled representation learning aims to isolate independent latent factors to improve model robustness and interpretability~\cite{0019CZM023,yao2024dual}. Early research utilized variational frameworks and information-theoretic constraints to separate macro-intentions from micro-attributes~\cite{0019CZM023}. More recently, multi-view extraction and self-supervised contrastive learning have been employed to capture multifaceted intents from heterogeneous relations and global collaborative signals~\cite{zhao2022multi,WangWHYLZG023,wang2022disencite,li2025disco}. While these approaches successfully identify what users prefer, they often struggle to distinguish why an interaction occurs in NBR settings—specifically, whether a purchase is driven by routine replenishment or novel exploration.
Existing NBR studies that address repeated behaviors typically perform disentanglement within a shared latent space~\cite{deng2023multi,wei2024knowledge}. However, this structural entanglement often allows dominant habit signals to overshadow subtle exploratory patterns, hindering interest discovery. {\model} adopts a gated dual-expert architecture to explicitly decouple these conflicting motives. An InfoNCE-based contrastive alignment task further enforces semantic coordination across the two expert spaces, ensuring the gating mechanism can coherently reconcile habit and exploration rather than arbitrarily blending entangled representations.

\section{Data Analysis}

The primary objective of this analysis is to investigate diverse behavioral characteristics across different consumption domains. By examining four real-world datasets—\textit{Beauty}, \textit{Sports}, \textit{Grocery}, and \textit{Home}—sourced from the Amazon benchmark~\cite{he2016ups}, we demonstrate that the balance between habit and exploration varies across categories, providing a statistical foundation and empirical motivation for our proposed dual-expert framework. The statistics of these datasets after pre-processing are summarized in Table~\ref{tab:dataINFO}. We focus on four key dimensions: repeat ratio evolution, temporal intervals, history length impact, and item-level repetition probability, as illustrated in Figure~\ref{fig_data}.

\noindent\textbf{Evolution of Repeat Ratio.} To understand the dynamics of user intent, the repeat ratio—defined as the fraction of re-purchased items in a basket—is tracked across ten chronological bins of user history (normalized by sequence length). As illustrated in Figure~\ref{fig_data}(a), evolutionary patterns vary significantly across domains. \textit{Grocery} exhibits a distinct inverted-U trend, rising as users establish consumption habits and declining later due to potential interest drift. In contrast, \textit{Beauty}, \textit{Sports}, and \textit{Home} maintain consistently lower repeat ratios, indicating a dominance of exploratory behavior. Specifically, \textit{Beauty} shows oscillatory patterns, while \textit{Home} remains relatively flat. These contrasting dynamics confirm that user intent is not static but evolves through different lifecycle stages, validating the necessity of a model that adaptively balances habit and exploration.

\noindent\textbf{Distribution of Repurchase Intervals.} To investigate item-specific purchase cycles, we analyze the repurchase interval, defined as the number of intervening baskets between two consecutive purchases of the same item ($\Delta t = pos_{current} - pos_{last}$). Figure~\ref{fig_data}(b) visualizes the probability density of these intervals using Kernel Density Estimation (KDE). As a non-parametric method, KDE provides a smooth and continuous estimation of the underlying probability distribution by mitigating the noise inherent in discrete histograms. This allows for a clearer identification of latent periodic modes—representing consistent replenishment cycles—and the specific rates of temporal decay across different categories. While most domains show a primary peak in the short term ($\Delta t < 10$), their long-term behaviors diverge significantly. Notably, \textit{Beauty} presents a unique bi-modal pattern where the second peak is nearly as intense as the initial one. These observations reveal that temporal influence is often non-monotonic, supporting our motivation to employ Hawkes-enhanced Fourier encoding rather than traditional static decay to capture these complex rhythms.

\noindent\textbf{Impact of History Length on Repurchase.} To determine how user maturity influences consumption habits, we analyze the correlation between user sequence length and the average repeat ratio. As shown in Figure~\ref{fig_data}(c), \textit{Grocery} exhibits a strong positive correlation, with the repeat ratio rising from $0.03$ to $\sim 0.08$ as history length increases. This confirms that habit strength solidifies over time in consumable domains. In \textit{Beauty}, a threshold effect is observed: the repeat ratio remains low until the sequence length exceeds 40, where it surges to $>0.04$. Conversely, \textit{Sports} and \textit{Home} show marginal growth regardless of history length. This suggests that for durable goods, user intent remains exploration-dominated, further justifying the need to decouple habit and exploration to avoid over-fitting to sparse repetitive signals.

\begin{figure*}[h]
	\centering
	\includegraphics[width=0.88\linewidth]{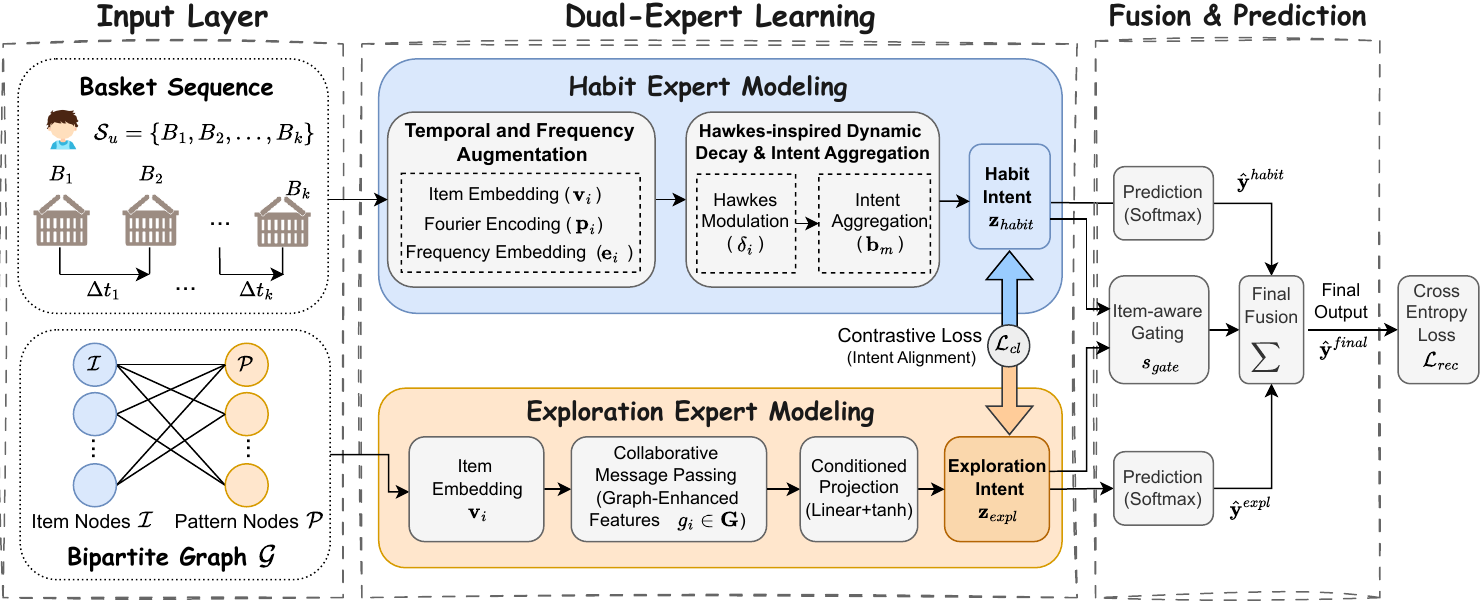}
	\caption{{\model} architecture, which consists of three key modules: (1) Habit Expert, which captures non-monotonic replenishment cycles via Hawkes-based intensity kernels and Fourier encoding; (2) Exploration Expert, which retrieves latent collaborative interests using an item-pattern bipartite graph; (3) Fusion and Prediction, which first bridges the semantic gap between the two experts through contrastive learning and adaptively fuses expert outputs via an item-aware gating mechanism.}
	\label{fig_model}
\end{figure*}

\noindent\textbf{Item-Level Repetition Probability.} To evaluate the repurchase tendency of items, we calculate the repetition probability, defined as the proportion of buyers who purchased an item more than once. Figure~\ref{fig_data}(d) presents the Cumulative Distribution Function of these probabilities. In \textit{Home} and \textit{Sports}, over 90\% of items have a repetition probability of zero, highlighting the severe sparsity in item recurrence for durable goods. In contrast, \textit{Grocery} exhibits a heavy-tailed distribution, with over 40\% of items eliciting repeat purchases. This disparity reinforces that robust exploration mechanisms are paramount for low-frequency categories (e.g., \textit{Home}) to capture latent interests amidst sparse repetitive signals. Furthermore, it underscores the significant behavioral heterogeneity across consumption domains, necessitating an adaptive modeling approach that can flexibly reconcile varying degrees of habit and exploration.

\section{Preliminary}
Let $\mathcal{U} = \{u_1, u_2, \dots, u_{|\mathcal{U}|}\}$ and $\mathcal{I} = \{i_1, i_2, \dots, i_{|\mathcal{I}|}\}$ denote the sets of users and items, respectively. For each user $u \in \mathcal{U}$, the interaction history is represented as a sequence of $k$ baskets:
\[\mathcal{S}_u = (B_1, B_2, \dots, B_k)\]
where $B_m \subseteq \mathcal{I}$ is the $m$-th basket in the chronological sequence.

\noindent\textbf{Task Definition.} Given the historical sequence $\mathcal{S}_u$, our objective is to predict the likelihood of each candidate item $i \in \mathcal{I}$ being included in the next basket $B_{k+1}$ and recommend a ranked list of Top-$K$ items. For simplicity, we omit the superscript $u$ in the following discussion when the context is clear.

\section{Methodology}

The overall architecture of {\model} is illustrated in Figure~\ref{fig_model}. Our framework follows a hierarchical pipeline that begins with an \textit{Input Layer} to transform basket sequences into dense embeddings, followed by a \textit{Dual-Expert Learning} module to capture distinct user motives, and concludes with a \textit{Fusion and Prediction} stage to generate the final recommendations. 

As shown in the framework, the core of our model lies in the parallel processing of user intents through two specialized components. First, the \textit{Habit Expert Modeling} (Section~\ref{1_HabitExpert}) is designed to characterize personalized, item-specific periodicities and dynamic intensity. By integrating Hawkes processes with Fourier time encoding~\cite{XiaoZTH25,CaiWQHZZ24}, this expert models the non-monotonic temporal rhythms of recurring behaviors to determine when historical items are likely to reappear. Simultaneously, the \textit{Exploration Expert} (Section~\ref{2_ExplorationExpert}) captures exploratory interests by leveraging a global item-pattern bipartite graph to mine potential item associations beyond the user's personal history.
Finally, the \textit{Fusion and Prediction} (Section~\ref{3_Prediction}) bridges the dual experts via an item-aware gating.

\subsection{Habit Expert Modeling} \label{1_HabitExpert} 
The Habit Expert is designed to capture recurring purchase patterns by modeling item-specific temporal dynamics. It consists of two primary stages: \textit{Hawkes-enhanced Temporal Modeling} and \textit{Habit Intent Aggregation}. The former integrates periodic rhythms and dynamic decay signals into item-level representations, while the latter summarizes these signals across the historical sequence to form a stable habit representation. 

\subsubsection{Hawkes-Enhanced Temporal Modeling}
This stage transforms static item embeddings into time-aware representations by augmenting the feature space with spectral periodicities and modulating them through an item-specific decay kernel.

\noindent\textbf{Temporal and Frequency Augmentation.}
Given an item embedding $\mathbf{v}_i \in \mathbb{R}^d$, we first integrate temporal periodicity and interaction intensity. Since our input consists of basket sequences, we define the relative temporal interval $\Delta t_i = (k+1) - m_{last}^{(i)}$, which denotes the number of baskets intervening between the target prediction and the item's most recent appearance at index $m_{last}^{(i)}$.
To characterize personalized periodicities, we employ a Fourier Time Encoding mechanism. This transformation maps the discrete interval $\Delta t_i$ into a periodic vector $\mathbf{p}_i \in \mathbb{R}^d$:
\begin{equation}
	\mathbf{p}_i = [\sin(\mathbf{w} \Delta t_i + \boldsymbol{\phi}); \cos(\mathbf{w} \Delta t_i + \boldsymbol{\phi})]
\end{equation}
where $\mathbf{w}, \boldsymbol{\phi} \in \mathbb{R}^{d/2}$ are learnable frequency and phase parameters. This spectral representation captures the non-monotonic behavioral rhythms identified in our empirical analysis. Additionally, let $\mathbf{e}_i \in \mathbb{R}^d$ be the frequency embedding corresponding to the item's historical purchase count. The enhanced item representation $\mathbf{v}'_i$ is formed as:
\begin{equation}\label{equ_vi}
	\mathbf{v}'_i = \mathbf{v}_i + w_t \mathbf{p}_i + w_f \mathbf{e}_i
\end{equation}
where $w_t$ and $w_f$ are learnable scalars that adaptively control the influence of temporal and frequency signals.

\noindent\textbf{Hawkes-Inspired Dynamic Decay.}
To model the diverse lifecycles of items, we introduce a decay mechanism inspired by the Hawkes process's intensity function. For each item $i$, we learn an item-specific decay rate $\lambda_i$ and calculate its dynamic influence $\delta_i$:
\begin{equation}\label{equ_delta}
	\lambda_i = \sigma(e_i^\lambda), \quad \delta_i = \exp(-\lambda_i \cdot \Delta t_i) + \epsilon
\end{equation}
where $e_i^\lambda \in \mathbb{R}^1$ is a learnable scalar and $\epsilon$ is a base intensity constant. This formulation reflects the excitation and subsequent decay of interest following an interaction. The final time-aware habit representation $\mathbf{h}_i$ is generated via element-wise modulation:
\begin{equation}
	\mathbf{h}_i = \delta_i \cdot \mathbf{v}'_i
\end{equation}
This ensures that the habit intent is conditioned on the dynamic excitation patterns inherent in the user's history.

\subsubsection{Habit Intent Aggregation}

The final habit intent $\mathbf{z}_{habit} \in \mathbb{R}^d$ is derived by aggregating the time-aware item representations $\mathbf{h}_i$ across the user's historical baskets. Given that a basket $B_m$ may contain items with varying relevance to a user's long-term habits, we employ an additive attention mechanism to dynamically determine the importance of each item. The attention weight $\beta_i$ for item $i \in B_m$ is calculated as:
\begin{equation}
	\beta_i = \frac{\exp\left(\mathbf{v}_{att}^\top \tanh(\mathbf{W}_{att} \mathbf{h}_i)\right)}{\sum_{j \in B_m} \exp\left(\mathbf{v}_{att}^\top \tanh(\mathbf{W}_{att} \mathbf{h}_j)\right)}
\end{equation}
where $\mathbf{W}_{att} \in \mathbb{R}^{d \times d}$ and $\mathbf{v}_{att} \in \mathbb{R}^d$ are learnable weight parameters. This mechanism ensures that items characterizing stable consumption patterns receive higher weights within the basket representation. Subsequently, the basket-level representation $\mathbf{b}_m$ and the comprehensive habit intent $\mathbf{z}_{habit}$ are computed as follows:
\begin{equation}
	\label{equ_bm}\mathbf{b}_m = \sum_{i \in B_m} \beta_i \mathbf{h}_i, \quad \mathbf{z}_{habit} = \tanh\left(\sum_{m=1}^{k} \mathbf{b}_m\right)
\end{equation}By enriching each item representation with spectral periodicities and purchase frequencies prior to this aggregation, the model effectively captures the varying lifecycles of diverse items while respecting their simultaneous occurrence within the same basket. This hierarchical design allows $\mathbf{z}_{habit}$ to function as a time-aligned summary of the user's recurring needs, providing a robust representation of habitual intent that remains distinct from the transient signals of exploratory behavior.

\subsection{Exploration Expert Modeling} \label{2_ExplorationExpert}
While the Habit Expert focuses on personal historical regularities, the Exploration Expert is designed to look outward to discover potential interests beyond the user's own history. To capture high-order collaborative signals and global item correlations, we propose a graph-based retrieval mechanism utilizing a global item-pattern bipartite graph. This expert complements the habit-based view by identifying items that align with broader consumption trends and shared purchasing behaviors across the entire community. This process is comprised of three primary modules: \textit{Global Item-Pattern Graph Construction} to establish the structural foundation, \textit{Collaborative Message Passing} to propagate global signals, and \textit{Conditioned Projection} to derive the final exploration intent.

\subsubsection{Global Item-Pattern Graph Construction}
To bridge the gap between individual interactions and global trends, we construct a bipartite graph $\mathcal{G} = (\mathcal{I} \cup \mathcal{P}, \mathcal{E})$, where $\mathcal{I}$ denotes the set of items and $\mathcal{P}$ represents a set of $M$ latent purchase patterns. Each pattern $p \in \mathcal{P}$ encapsulates a group of items that frequently co-occur across different users' baskets, representing latent functional or behavioral clusters. An edge $(i, p) \in \mathcal{E}$ exists if item $i$ is associated with pattern $p$. By projecting items into this shared pattern space, the model transcends the limitations of a single user's history and taps into the collective intelligence of the dataset to identify relevant but previously unpurchased items.

\subsubsection{Collaborative Message Passing}
To inject global collaborative signals into the item representations, we perform explicit message passing on $\mathcal{G}$ using a Graph Encoder. Inspired by the simplified propagation rules in LightGCN~\cite{LightGCN}, we adopt a lightweight linear aggregation scheme to ensure efficient propagation of high-order correlations. This process updates item embeddings by traversing the latent pattern nodes in two distinct stages. First, we compute the representation of each pattern node $\mathbf{e}_p \in \mathbb{R}^d$ by aggregating the embeddings of its constituent items. We employ symmetric normalization to account for item popularity and pattern scale:
\begin{equation}\mathbf{e}_p = \sum_{i \in \mathcal{N}_p} \frac{1}{\sqrt{|\mathcal{N}_p| |\mathcal{N}_i|}} \mathbf{v}_i\end{equation}
where $\mathcal{N}_p$ denotes the set of items connected to pattern $p$, and $|\mathcal{N}_p|, |\mathcal{N}_i|$ represent the degrees of the pattern node and item node, respectively. Subsequently, we update the item embeddings to reflect their global context by propagating the pattern signals back to the item nodes. To preserve the item's inherent identity while incorporating collaborative context, the final graph-enhanced embedding $\mathbf{g}_i \in \mathbb{R}^d$ is formed as:
\begin{equation}\label{equ_gi}
	\mathbf{g}_i = \mathbf{v}_i + \sum_{p \in \mathcal{N}_i} \frac{1}{\sqrt{|\mathcal{N}_i| |\mathcal{N}_p|}} \mathbf{e}_p\end{equation}
where $\mathcal{N}_i$ is the set of patterns associated with item $i$. The resulting representation $\mathbf{g}_i$ captures not only the item's static semantics but also its collaborative role within global consumption patterns.

\subsubsection{Conditioned Projection}
Exploration in our framework is not a random discovery process but is adaptively guided by the user’s current habitual state. We derive the exploration intent vector $\mathbf{z}_{expl} \in \mathbb{R}^d$ by projecting the summarized habit representation $\mathbf{z}_{habit}$ into the exploration latent space:
\begin{equation}
	\mathbf{z}_{expl} = \tanh(\mathbf{W}_{expl} \mathbf{z}_{habit} + \mathbf{b}_{expl})
\end{equation}
where $\mathbf{W}_{expl} \in \mathbb{R}^{d \times d}$ and $\mathbf{b}_{expl} \in \mathbb{R}^d$ are learnable projection parameters. This vector encodes the user’s potential interests conditioned on their existing habits. By mapping the stable habit signal into the exploration space, the model generates a search query that identifies items aligned with high-order collaborative patterns (represented by $\mathbf{g}_i$). This architecture ensures that the discovery of new items is grounded in global item-item correlations while remaining sensitive to the user’s specific behavioral trajectory, preventing the exploration expert from drifting toward generic popularity signals. The resulting $\mathbf{z}_{expl}$ serves as the final representation for the exploration expert.

$\mathbf{z}_{habit}$ is derived from the user’s personal interaction history, while $\mathbf{z}_{expl}$ is scored against the global graph embedding matrix $\mathbf{G}$, which explicitly encodes community-wide co-occurrence patterns. These heterogeneous objectives pull the two representations into functionally distinct subspaces: $\mathbf{z}_{habit}$ serves as a \textit{personalized query} rooted in the user’s identity, while $\mathbf{z}_{expl}$ acts as its rotated counterpart aligned with the complementary direction most informative for retrieving novel items from the item-pattern graph.

\subsection{Intent Fusion and Prediction}
\label{3_Prediction}

This stage dynamically integrates the Habit and Exploration experts through an item-aware gating mechanism, ensuring that the final recommendation accounts for both historical regularities and global collaborative signals. This process comprises three primary modules: \textit{Item-Aware Gating} to balance user motives at a fine-grained level, \textit{Fusion and Prediction} to ensure semantic consistency via contrastive learning, and to generate the final ranked list.

\subsubsection{Item-Aware Gating Mechanism}Traditional NBR methods often employ a global fusion strategy that applies a uniform weight to all items in a basket. However, empirical observations suggest that a single basket often reflects a mixture of conflicting motives, such as the routine purchase of \textit{Milk} (Habit) alongside the exploratory trial of \textit{Cola} (Exploration).
To capture this fine-grained distinction, we propose an item-aware gating mechanism that aims to decouple these entangled user intents at the item level. Specifically, for each candidate item $i$, the mechanism determines whether it aligns more closely with the user's habitual trajectory or their potential exploratory interest. We first derive a unified gating context $\mathbf{s}_{gate} \in \mathbb{R}^d$ by projecting the concatenated dual-intent vectors into a shared hidden space:\begin{equation}\mathbf{s}_{gate} = \tanh\left( \mathbf{W}_{f} [\mathbf{z}_{habit} \oplus \mathbf{z}_{expl}] + \mathbf{b}_{f} \right)
\end{equation}
where $\oplus$ denotes concatenation, and $\mathbf{W}_{f} \in \mathbb{R}^{d \times 2d}, \mathbf{b}_{f} \in \mathbb{R}^d$ are learnable parameters. The item-specific gate value $\alpha_i$ is then computed by matching this gating context with the target item's static embedding $\mathbf{v}_i$:
\begin{equation}
	\alpha_i = \sigma\left( \mathbf{s}_{gate}^\top \mathbf{v}_i \right)\end{equation}where $\alpha_i \in (0,1)$ is the item-aware gate value derived from the fused dual-intent context. In this formulation, $\alpha_i \to 1$ indicates that the candidate is a habitual response, thus prioritizing the Habit Expert, while $\alpha_i \to 0$ suggests the item is a discovery driven by global item-pattern correlations, favoring the Exploration Expert. This per-item strategy allows the model to flexibly transition between consistency and novelty for different candidates within the same prediction window.

\subsubsection{Fusion and Prediction}
The final recommendation is generated by ensembling the probability distributions from both experts, integrating high-order collaborative signals with the user's sequential history. The computational flow for deriving the final preference scores is defined as follows:
\begin{equation}
	\left\{
	\begin{aligned}
		\hat{\mathbf{y}}^{habit} &= \text{Softmax}(\mathbf{W}_{out} \mathbf{z}_{habit}) \\
		\hat{\mathbf{y}}^{expl} &= \text{Softmax}(\mathbf{z}_{expl} \mathbf{G}^\top) \\
		\hat{\mathbf{y}}^{final} &= \bm{\alpha} \odot \hat{\mathbf{y}}^{habit} + (\mathbf{1} - \bm{\alpha}) \odot \hat{\mathbf{y}}^{expl}
	\end{aligned}
	\right.
\end{equation}
where $\mathbf{W}_{out} \in \mathbb{R}^{|\mathcal{I}| \times d}$ is the output mapping matrix.
$\mathbf{G} \in \mathbb{R}^{|\mathcal{I}| \times d}$ represents the global item embedding matrix, where each row $i$ corresponds to the graph-enhanced representation $\mathbf{g}_i$ derived in Equation~(\ref{equ_gi}).
The vector $\bm{\alpha} = [\alpha_1, \alpha_2, \dots, \alpha_{|\mathcal{I}|}]$ contains the item-aware gating scores, which allow the model to flexibly transition between exploitation and exploration at a fine-grained, item-centric level.
By combining these distributions, the model ensures that the final recommendation list is sensitive to both the user's specific replenishment cycles and broader items of interest discovered through collaborative patterns.

\subsection{Optimization Objective}
We optimize the dual-expert architecture through a multi-task objective that jointly optimizes recommendation accuracy and representation alignment. The primary task minimizes the cross-entropy loss between the fused prediction $\hat{\mathbf{y}}^{final}_{ui}$ and the ground-truth basket $B_{k+1}$:
\begin{equation}
	\mathcal{L}_{rec} = - \sum_{u \in \mathcal{U}} \sum_{i \in B_{k+1}} \log(\hat{y}_{u,i}^{final})
\end{equation}
To ensure both experts converge on a unified user representation, we introduce an InfoNCE-based contrastive alignment loss. By treating $(\mathbf{z}_{habit}^{(i)}, \mathbf{z}_{expl}^{(i)})$ as positive pairs and mismatched users within a mini-batch of size $N$ as negatives, the loss is defined as:
\begin{equation}
	\mathcal{L}_{cl} = - \sum_{i=1}^N \log \frac{\exp(\text{sim}(\mathbf{z}_{habit}^{(i)}, \mathbf{z}_{expl}^{(i)}) / \tau)}{\sum_{j=1}^N \exp(\text{sim}(\mathbf{z}_{habit}^{(i)}, \mathbf{z}_{expl}^{(j)}) / \tau)}
\end{equation}
where $\tau$ is the temperature and $\text{sim}(\cdot)$ is cosine similarity. The total joint objective $\mathcal{L}_{total}$ integrates these tasks with $L_2$ regularization to prevent overfitting:
\begin{equation}
	\mathcal{L}_{total} = \mathcal{L}_{rec} + \mu \mathcal{L}_{cl} + \gamma \|\Theta\|_2^2
\end{equation}
where $\mu$ and $\gamma$ are hyperparameters controlling the influence of the contrastive task and weight decay for all learnable parameters $\Theta$. This objective is optimized by Adam~\cite{Adam_2015}.

\section{Experiments}
\label{sec:experiment}

\setlength{\tabcolsep}{5pt}
\begin{table*}[t]
	\centering
	\caption{Performance comparison of {\model} against representative baselines on four datasets. The second best results are underlined, while the overall best results are highlighted in bold. Improvements achieved by {\model} are statistically significant based on the paired t-test ($p<0.05$). The last three rows report the ablation study results of different model components.}
	\label{tab_overall_ablation}
	\resizebox{\linewidth}{!}{
		\begin{tabular}{lcccccccccccccccc}
			\toprule
			\multirow{2}{*}{Method} 	& \multicolumn{4}{c}{Beauty} 
			& \multicolumn{4}{c}{Sports} 
			& \multicolumn{4}{c}{Grocery} 
			& \multicolumn{4}{c}{Home} \\
			\cmidrule(lr){2-5} \cmidrule(lr){6-9} \cmidrule(lr){10-13} \cmidrule(lr){14-17}
			
			& R@10 & N@10 & R@20 & N@20
			& R@10 & N@10 & R@20 & N@20
			& R@10 & N@10 & R@20 & N@20
			& R@10 & N@10 & R@20 & N@20 \\
			\midrule
			
			POP
			& 0.0092 & 0.0038 & 0.0167 & 0.0054
			& 0.0078 & 0.0091 & 0.0093 & 0.0118
			& 0.0165 & 0.0074 & 0.0197 & 0.0095
			& 0.0038 & 0.0056 & 0.0042 & 0.0066 \\
			
			FPMC
			& 0.0101 & 0.0043 & 0.0177 & 0.0068
			& 0.0032 & 0.0015 & 0.0045 & 0.0019
			& 0.0198 & 0.0089 & 0.0218 & 0.0096
			& 0.0046 & 0.0052 & 0.0057 & 0.0067 \\
			
			Sets2Sets
			& 0.0628 & 0.0416 & 0.1271 & 0.0525
			& 0.0165 & 0.0229 & 0.0378 & 0.0334
			& 0.0487 & 0.0295 & 0.0618 & 0.0328
			& 0.0053 & 0.0057 & 0.0076 & 0.0071 \\
			
			TIFU-KNN
			& 0.0555 & 0.0282 & 0.0851 & 0.0373
			& 0.0125 & 0.0055 & 0.0216 & 0.0081
			& 0.0556 & 0.0333 & 0.0932 & 0.0445
			& 0.0056 & 0.0066 & 0.0082 & 0.0073 \\
			%			\midrule
			
			MMNR
			& 0.0774 & 0.0371 & 0.1338 & 0.0499
			& 0.0288 & 0.0251 & 0.0447 & 0.0293
			& 0.0373 & 0.0271 & 0.0584 & 0.0352
			& 0.0076 & 0.0082 & 0.0158 & 0.0093 \\
			
			%			\midrule
			M$^2$
			& 0.1077 & \underline{0.0639} & 0.1497 &\underline{ 0.0811}
			& 0.0256 & 0.0288 & 0.0438 & \underline{0.0334}
			& 0.0445 & 0.0314 & 0.0654 & 0.0384
			& 0.0092 & 0.0098 & 0.0165 & \underline{0.0124} \\
			
			MCRec
			& 0.0931 & 0.0499 & 0.1554 & 0.0669
			& 0.0316 & 0.0259 & 0.0501 & 0.0313
			& \underline{0.0583} & 0.0406 & 0.0839 & 0.0491
			& 0.0097 & 0.0079 & 0.0176 & 0.0091 \\
			
			HapCL
			& 0.0915 & 0.0428 & 0.1549 & 0.0648
			& 0.0308 & 0.0252 & 0.0519 & 0.0318
			& 0.0544 & 0.0384 & 0.0851 & 0.0477
			& 0.0094 & 0.0077 & 0.0171 & 0.0085 \\
			
			SemTHy
			& \underline{0.1078} & 0.0597 & \underline{0.1565} & 0.0784
			& \underline{0.0369} & 0.0274 & \underline{0.0532} & 0.0327
			& 0.0566 & 0.0394 & 0.0867 & 0.0485
			& \underline{0.0131} & 0.0089 & \underline{0.0212} & 0.0117 \\
			
			TREx
			& 0.0984 & 0.0626 & 0.1418 & 0.0754
			& 0.0350 & \underline{0.0296} & 0.0408 & 0.0310
			& 0.0577 & \underline{0.0407} & \underline{0.1027} & \underline{0.0550}
			& 0.0123 & \underline{0.0099} & 0.0168 & 0.0113 \\
			
			\midrule
			\textbf{TIDE}
			& \textbf{0.1276} & \textbf{0.0728} & \textbf{0.2076} & \textbf{0.0951}
			& \textbf{0.0434} & \textbf{0.0308} & \textbf{0.0560} & \textbf{0.0351}
			& \textbf{0.0675} & \textbf{0.0448} & \textbf{0.1168} & \textbf{0.0605}
			& \textbf{0.0160} & \textbf{0.0102} & \textbf{0.0286} & \textbf{0.0142} \\
			
			\midrule
			w/o Time
			& 0.1112 & 0.0594 & 0.1822 & 0.0795
			& 0.0339 & 0.0219 & 0.0554 & 0.0291
			& 0.0644 & 0.0447 & 0.1157 & 0.0524
			& 0.0153 & 0.0094 & 0.0279 & 0.0135 \\
			
			w/o Graph
			& 0.0510 & 0.0257 & 0.0855 & 0.0350
			& 0.0181 & 0.0175 & 0.0327 & 0.0218
			& 0.0308 & 0.0190 & 0.0439 & 0.0232
			& 0.0017 & 0.0007 & 0.0028 & 0.0011 \\
			
			w/o CL
			& 0.1153 & 0.0662 & 0.1989 & 0.0891
			& 0.0421 & 0.0300 & 0.0517 & 0.0341
			& 0.0648 & 0.0427 & 0.1117 & 0.0582
			& 0.0148 & 0.0096 & 0.0284 & 0.0134 \\
			
			\bottomrule
		\end{tabular}
	}
\end{table*}

We conduct experiments to evaluate the effectiveness of \model. We first describe the experimental setup and then present the experimental results. Additionally, we perform ablation study to verify the effectiveness of each module, analyze the impact of different sequence encoders, and finally discuss the parameter sensitivity.

\subsection{Experimental Setup}
\noindent\textbf{Datasets.}
We evaluate {\model} on four datasets from the Amazon benchmark~\cite{he2016ups}: \textit{Grocery} (high-frequency, habitual), \textit{Home} and \textit{Sports} (low-frequency, exploration-driven), and \textit{Beauty} (hybrid, oscillatory). These datasets provide a diverse range of sparsity levels and behavioral patterns, as summarized in Table~\ref{tab:dataINFO}.
Following established protocols~\cite{TIFUKNN_2020,CLEA_2021,PengM22023}, we apply 5-core filtering to remove infrequent users and items. Interactions are grouped into chronological basket sequences based on transaction timestamps. For model evaluation, we adopt the leave-one-out strategy~\cite{SRs_hier,Caser,PengM22023}: the final basket of each user is reserved for testing, the penultimate for validation, and the remainder for training. This setup rigorously tests the model’s ability to reconcile historical habits with future exploratory intents.

\noindent\textbf{Baselines.}
We compare the proposed {\model} against the following representative state-of-the-art NBR methods:
\begin{itemize}[leftmargin=*, nosep, font=\textbf]
	\item {\textbf{POP}} ranks items based on their overall popularity across all users, and recommends the Top-$K$ popular items.
	\item {\textbf{FPMC}}~\cite{FPMC_2010} is a shallow model that integrates hidden factors and first-order Markov chain for next-basket recommendation.
	\item {\textbf{Sets2Sets}}
	~\cite{HuSets2Sets2019}
	is an end-to-end method for multiple baskets prediction based on RNN. It also incorporates the repeated purchase pattern into the model.
	\item {\textbf{TIFU-KNN}}
	~\cite{TIFUKNN_2020} is a $k$-nearest neighbors method that generates recommendations based on the historical behavior of the target user and its similarity to other users.
	\item {\textbf{MMNR}}
	~\cite{deng2023multi}: A multi-view multi-aspect model that removes spurious interests and captures low-level item correlations via disentangled item representations.
	\item \textbf{M}$^2$	
	~\cite{PengM22023} 
	explicitly models three significant factors in the next-basket generation process: users' general preferences, items' popularities, and transition patterns among items.
	\item {\textbf{MCRec}}
	~\cite{deng2024multi}: A multi-scale context-aware model that captures diverse behavioral patterns via tensorized basket modeling and adaptive interest fusion.
	\item {\textbf{HapCL}
		~\cite{su2024hierarchical}}: Mines information from multiple views and patterns with hierarchical modeling, and introduces polar contrastive learning with graph-based augmentation to alleviate data sparsity and enhance representation quality.
	\item {\textbf{SemTHy}
		~\cite{zhou2025dual}}: A dual-tower model that fuses semantic and collaborative relations into user and item representations via semantic perception and a timespan-coupled hypergraph.
	\item {\textbf{TREx}}
	~\cite{li2024we}: A plug-and-play repetition–exploration framework that ensures high accuracy while optimizing beyond-accuracy objectives.
\end{itemize}

\noindent\textbf{Evaluation Metrics.}
To evaluate the performance of {\model}, we employ two widely used metrics in the NBR task~\cite{TIFUKNN_2020,PengM22023,MBN_2022}: Recall@$K$ (R@$K$) and Normalized Discounted Cumulative Gain@$K$ (N@$K$). R@$K$ measures the proportion of ground-truth items successfully retrieved within the Top-$K$ recommendations, while N@$K$ accounts for the specific ranking positions of these items. Higher values across both metrics indicate superior recommendation quality. In our experiments, we report results for $K \in \{10, 20\}$.

\vspace{0.5em}
\noindent\textbf{Implementation Details.}
To ensure a fair and rigorous comparison, all models are implemented under a consistent experimental framework. We employ the Adam optimizer with a batch size of 100. The embedding size and hidden layer dimension $d$ are both fixed at 32. For all models, the learning rate and $L_2$ regularization weight are tuned via grid search within the range $\{10^{-4}, 10^{-3}, 10^{-2}, 10^{-1}\}$. For our proposed {\model}, we initialize the base intensity constant $\epsilon$ and the learnable decay parameter $\lambda$ to 0.1. For the contrastive alignment task, the temperature parameter $\tau$ is set to 0.1. To ensure the reliability of our findings, we perform paired $t$-tests across all experiments to assess statistical significance. All other hyperparameters for baseline models are tuned to their optimal values as reported in their original papers or determined through empirical search.
Our codes can be found at \url{https://github.com/Hiiizhy/TIDE}.

\subsection{Performance Comparison}

Table~\ref{tab_overall_ablation} summarizes the performance of {\model} against various representative baselines across four diverse datasets. Several key observations can be made regarding its effectiveness. First and foremost, {\model} consistently achieves the overall best performance across all datasets and evaluation metrics, demonstrating its robustness in handling different consumption behaviors. For instance, in the \textit{Beauty} dataset, {\model} substantially outperforms the strongest baseline, SemTHy, by 18.4\% in R@10 and 21.9\% in N@10. Similarly, in the \textit{Grocery} domain—which is characterized by high-frequency repurchase habits—our framework reaches a Recall@20 of 0.1168, marking a significant improvement over competitive models like TREx (0.1027) and MCRec (0.0839). These consistent gains across both high-frequency consumables and low-frequency durable goods validate that {\model} effectively reconciles habitual replenishment with exploratory discovery regardless of the underlying data sparsity.

The failure of traditional methods such as POP and FPMC underscores that simple popularity or first-order Markov transitions are insufficient to model the complex dynamics of basket sequences. While deep learning models like Sets2Sets and TIFUKNN specifically target repetition patterns, their performance is inherently limited by their inability to handle non-monotonic temporal influences. As revealed in our earlier empirical analysis, repurchase intervals often exhibit complex bi-modal structures; methods relying on static averages or simple RNNs fail to adapt to these fluctuations. In contrast, the Hawkes-enhanced Habit Expert in {\model} utilizes intensity jumps to reset and maintain high repurchase probabilities at the precise moment of subsequent interactions, capturing periodic rhythms that other baselines overlook.

Furthermore, while competitive baselines such as M$^2$, TREx, and SemTHy provide stronger results by incorporating global transitions or semantic information, they typically suffer from entangled representations where habitual and exploratory motives are conflated into a single vector. This entanglement often leads to a recommendation bias toward popular items, especially in the \textit{Home} and \textit{Sports} datasets where over 90\% of items have zero repetition probability. {\model} addresses this by using the habit state as an anchor to explore a global item-pattern graph. This decoupled approach ensures that exploratory recommendations remain semantically aligned with the user’s core interests, allowing {\model} to maintain high accuracy even in exploration-dominated environments. By explicitly disentangling habit from exploration and enhancing temporal modeling through a Hawkes-inspired kernel, our framework avoids the noise often introduced by neighbor-based augmentation while capturing the diverse and intricate dynamics of modern consumption patterns.

\subsection{Ablation Study}
In this section, we conduct ablation experiments to evaluate the contribution of each key component in {\model}. The results, presented in the final three rows of Table~\ref{tab_overall_ablation}, compare the full model against three variants: w/o Time (removing Hawkes-enhanced Temporal Modeling), w/o Graph (removing the global graph-based Exploration Expert), and w/o CL (removing the Contrastive Alignment task). The most substantial performance degradation occurs in the w/o Graph variant, where the absence of the Pattern-Guided Exploration Expert causes Recall@10 to collapse by 89.4\% in the \textit{Home} dataset (from 0.0160 to 0.0017) and by 60.0\% in \textit{Beauty} (from 0.1276 to 0.0510). This drastic decline confirms that relying solely on personal history is insufficient for next-basket recommendation and underscores the necessity of leveraging global collaborative signals via our item-pattern bipartite graph to discover latent interests, particularly in sparse domains where exploratory behavior dominates.
Furthermore, the w/o Time variant validates the importance of modeling item-specific temporal dynamics. Removing the Hawkes-enhanced module leads to a noticeable drop in performance across all domains, particularly in the \textit{Beauty} and \textit{Grocery} datasets where our empirical analysis identified strong periodic rhythms. The lack of \textit{intensity jump} modeling in this variant prevents the model from accurately resetting repurchase probabilities, confirming that static position embeddings are inadequate for capturing the complex, non-monotonic replenishment cycles inherent in consumer behavior. Finally, the w/o CL variant consistently performs worse than the full model, as seen in the \textit{Sports} dataset where Recall@10 decreases from 0.0434 to 0.0421. While the two experts provide the primary predictive power, the InfoNCE-based contrastive task is essential for bridging the semantic gap between the sequential (habit) and graph-based (exploration) latent spaces. By aligning these dual intents, {\model} achieves a more coherent fusion and reduces the noise introduced by disparate feature sources, leading to more stable and accurate recommendations.

\begin{figure}[t]
	\centering
	\includegraphics[width=0.85\linewidth]{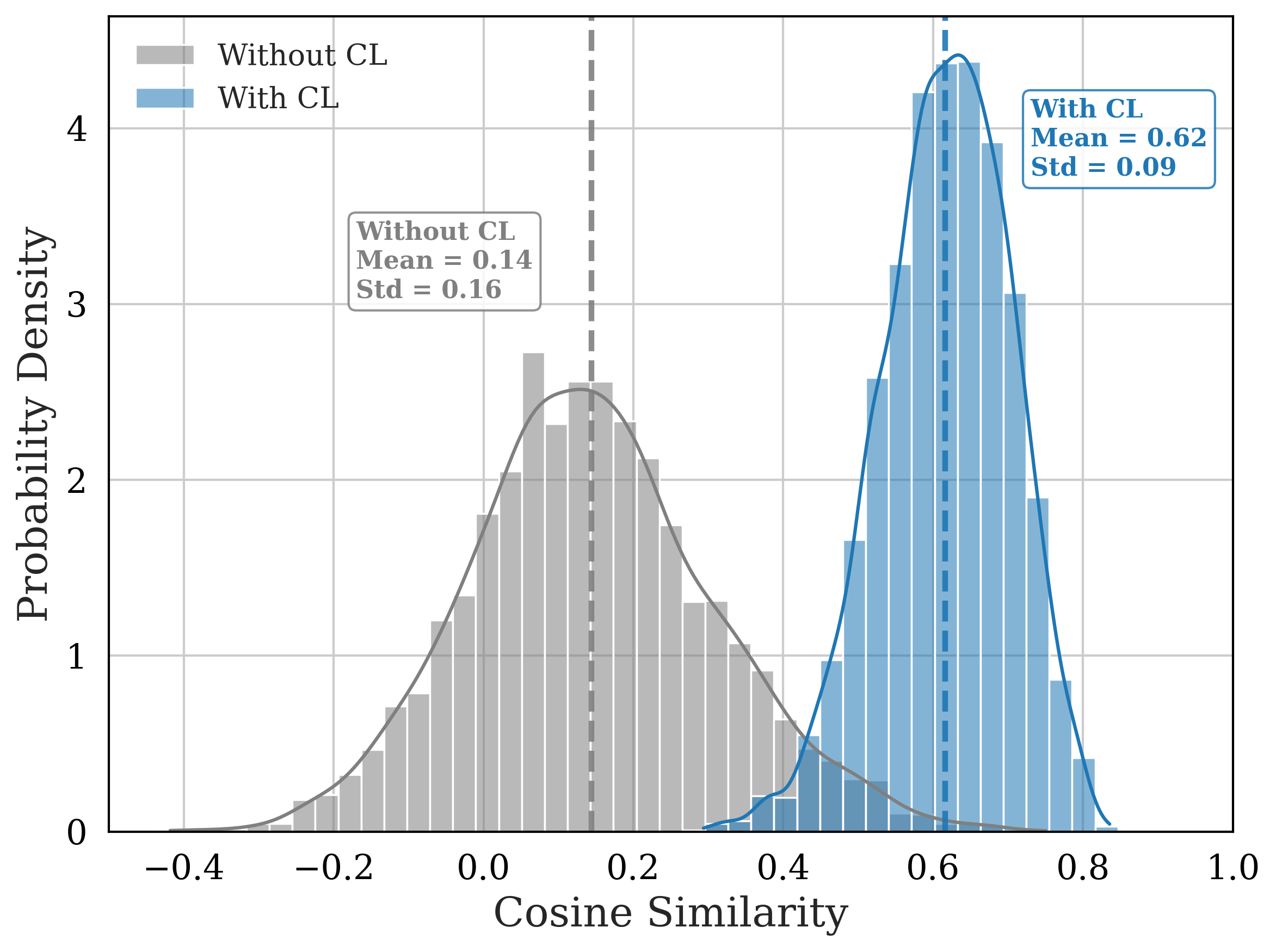}
	\caption{Probability density distribution of semantic alignment between dual-expert intents. The x-axis denotes the cosine similarity between the habit intent representation ($\mathbf{z}_{habit}$) and the exploration intent representation ($\mathbf{z}_{expl}$), while the y-axis represents the probability density.}
	\label{fig_align}
\end{figure}

\begin{figure*}[t]
	\centering
	\includegraphics[width=\linewidth]{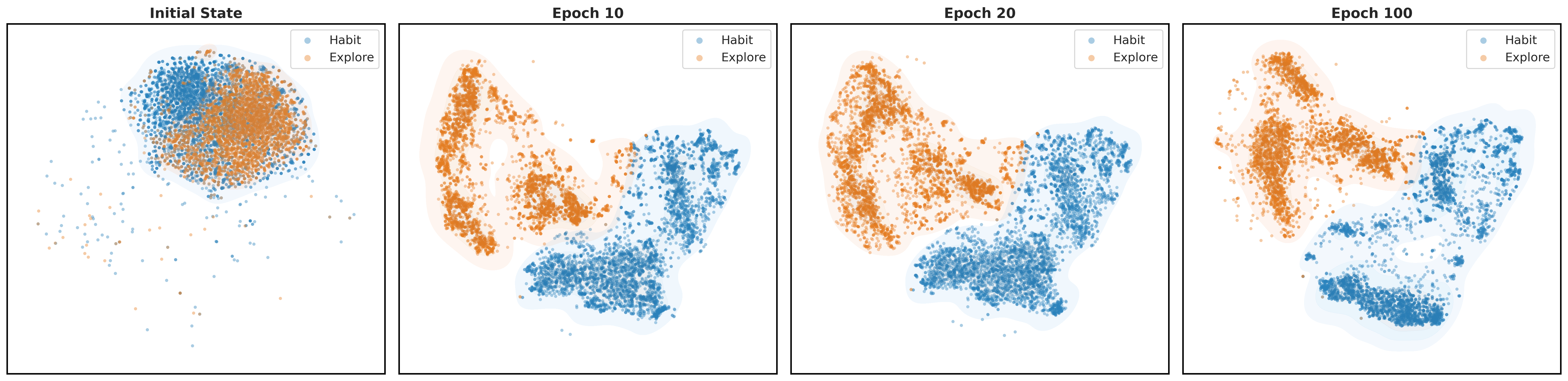}
	\caption{t-SNE visualization of the evolution of habit intent ($\mathbf{z}_{habit}$) and exploration intent ($\mathbf{z}_{expl}$) across different training stages. The transition from a chaotic, entangled distribution at the initial state to distinct yet semantically aligned clusters at Epoch 100 demonstrates the effective disentanglement and coordinated specialization achieved by the dual-expert framework.}
	\label{fig_tsne_evolution}
\end{figure*}

\subsection{Analysis of Intent Alignment}
Figure~\ref{fig_align} provides a quantitative look at how the contrastive alignment task bridges the semantic gap between our dual experts. By calculating the cosine similarity between the habit intent ($\mathbf{z}_{habit}$) and exploration intent ($\mathbf{z}_{expl}$), we observe a stark contrast in semantic coordination. The distribution for the \textit{Without CL} variant is notably dispersed, spanning a wide range from -0.4 to 0.8. This broad spread indicates that without explicit alignment, the two experts often produce divergent or contradictory representations, forcing the gating mechanism to synthesize predictions from semantically disjoint vectors. In contrast, the \textit{With CL} variant exhibits a highly concentrated and significantly right-shifted distribution, with most scores falling between 0.3 and 0.8. The elimination of negative correlations and the higher mean similarity prove that our InfoNCE objective successfully anchors the exploration intent to the habitual base. This forced alignment ensures a consistent semantic foundation, allowing the gating mechanism to reconcile the two motives more effectively and explaining the performance gains noted in our main results.

\subsection{Impact of Pattern Mining Strategies}
We evaluate the robustness of the Exploration Expert by replacing the default co-occurrence count used in {\model} with NPMI, Jaccard Similarity, and Lift when constructing the bipartite graph $\mathcal{G}$. As summarized in Table~\ref{tab:pattern_analysis}, {\model} remains robust across all strategies, though the optimal choice aligns closely with specific domain behaviors. On the \textit{Beauty} dataset, {\model}$_{\text{Jaccard}}$ achieves the highest performance, likely because normalizing by the purchaser union effectively uncovers complementary item sets while suppressing popularity bias. For dense, high-frequency domains like \textit{Grocery}, the default {\model} is superior, as absolute co-occurrence frequency serves as a highly reliable signal when interactions are abundant. In contrast, the sparse \textit{Home} dataset benefits most from {\model}$_{\text{Lift}}$ and {\model}$_{\text{NPMI}}$, which prioritize significant associations that exceed random chance. These findings confirm that while {\model} provides a stable foundation across various strategies, tailoring the mining metric to a domain's specific sparsity and repurchase rhythm can further refine its ability to capture latent collaborative interests.

\subsection{Impact of Different Sequential Modeling Strategies}

This section investigates whether actual temporal intervals ($\Delta t$) or simple sequential order provides a more effective signal for next-basket recommendation. We compare {\model} against two variants: {\model}$_{\text{Pos}}$, which uses static position embeddings, and {\model}$_{\text{GRU}}$, which employs a recurrent unit to capture discrete sequential dependencies. The experimental results demonstrate the clear superiority of time-aware modeling over simple order-based methods. {\model} consistently outperforms both variants, showing its most significant gains in sparse datasets. This confirms that elapsed time is a far more informative signal than relative positioning; while position embeddings or recurrent units treat transitions as uniform steps, our Hawkes-enhanced model uses item-specific decay rates to distinguish between short-term and long-term gaps. Although {\model}$_{\text{GRU}}$ remains competitive in highly consistent domains like grocery shopping, its performance degrades in sparse environments. By grounding the habit expert in the physical timeline and capturing intensity jumps through a Hawkes-inspired kernel, {\model} accurately models the timing of repurchases that traditional recurrent or static methods inherently overlook. The results are shown in Table~\ref{tab_pattern}.

\setlength{\tabcolsep}{5pt}
\begin{table}[t]
	\centering
	\caption{Performance on different Pattern Mining Strategies.}
	\vspace{-0.5em}
	\label{tab:pattern_analysis}
	\resizebox{\linewidth}{!}{
		\begin{tabular}{llcccc}
			\hline
			{Dataset} & {Metric} & {\model}${_{\text{NPMI}}}$ & {\model}${_{\text{Jaccard}}}$ & {\model}${_{\text{Lift}}}$ & {\model} \\
			\midrule
			\multirow{2}{*}{Beauty}
			& Recall@10 & 0.1219 & 0.1331 & 0.1227 & 0.1276 \\
			& NDCG@10 & 0.0669 & 0.0716 & 0.0677 & 0.0728 \\
			\hline
			\multirow{2}{*}{Sports}
			& Recall@10 & 0.0425 & 0.0412 & 0.0423 & 0.0434 \\
			& NDCG@10 & 0.0322 & 0.0289 & 0.0281 & 0.0308 \\
			\hline
			\multirow{2}{*}{Grocery}
			& Recall@10 & 0.0666 & 0.0641 & 0.0653 & 0.0675 \\
			& NDCG@10 & 0.0469 & 0.0437 & 0.0461 & 0.0448 \\
			\hline
			\multirow{2}{*}{Home}
			& Recall@10 & 0.0170 & 0.0168 & 0.0171 & 0.0160 \\
			& NDCG@10 & 0.0102 & 0.0098 & 0.0100 & 0.0102 \\
			\hline
		\end{tabular}
	}
\end{table}

\setlength{\tabcolsep}{8pt}
\begin{table}[t]
	\centering
	\caption{Effect of Different Sequential Modeling Strategies.}
	\vspace{-0.5em}
	\label{tab_pattern}
	\resizebox{\linewidth}{!}{
		\begin{tabular}{llccc}
			\hline
			{Dataset} & {Metric} & {\model}${_{\text{Pos}}}$ & {\model}${_{\text{GRU}}}$ & {\model} \\
			\hline
			\multirow{2}{*}{Beauty}
			& Recall@10 & 0.1055 & 0.1117 & {0.1276} \\
			& NDCG@10 & 0.0559 & 0.0619 & {0.0728} \\
			\hline
			\multirow{2}{*}{Sports}
			& Recall@10 & 0.0333 & 0.0358 & {0.0434} \\
			& NDCG@10 & 0.0230 & 0.0282 & {0.0308} \\
			\hline
			\multirow{2}{*}{Grocery}
			& Recall@10 & 0.0473 & 0.0672 & {0.0675} \\
			& NDCG@10 & 0.0301 & 0.0442 & {0.0448} \\
			\hline
			\multirow{2}{*}{Home}
			& Recall@10 & 0.0055 & 0.0150 & {0.0160} \\
			& NDCG@10 & 0.0047 & 0.0099 & {0.0102} \\
			\hline
		\end{tabular}
	}
\end{table}

\subsection{Visualization of Latent Representation}
Figure~\ref{fig_tsne_evolution} illustrates the evolution of the latent representation space using t-SNE~\cite{maaten2008visualizing}, revealing how {\model} learns to distinguish and align user motives. By projecting the habit intent ($\mathbf{z}_{habit}$) and exploration intent ($\mathbf{z}_{expl}$) from the initial state to Epoch 100, we observe a clear trajectory toward coordinated specialization. Initially, representations are heavily entangled and randomly distributed. As training progresses, a clear trend of disentanglement emerges; by Epoch 100, the vectors form concentrated clusters, indicating that the model has identified consistent behavioral prototypes for both motives. Notably, these clusters maintain strategic proximity and partial overlap. This proximity is evidence of successful Intent Alignment: the relative closeness ensures that both experts \textit{speak the same semantic language}, allowing the gating mechanism to smoothly transition between replenishment and discovery within a shared context. The transition from chaotic entanglement to a structured, dual-cluster formation confirms that our multi-task objective effectively balances the recommendation loss ($\mathcal{L}_{rec}$) and contrastive loss ($\mathcal{L}_{cl}$), guiding the system toward a state of balanced specialization.

\subsection{Parameter Sensitivity Study}
\begin{figure}[h]
	\centering
	\includegraphics[width=0.46\linewidth]{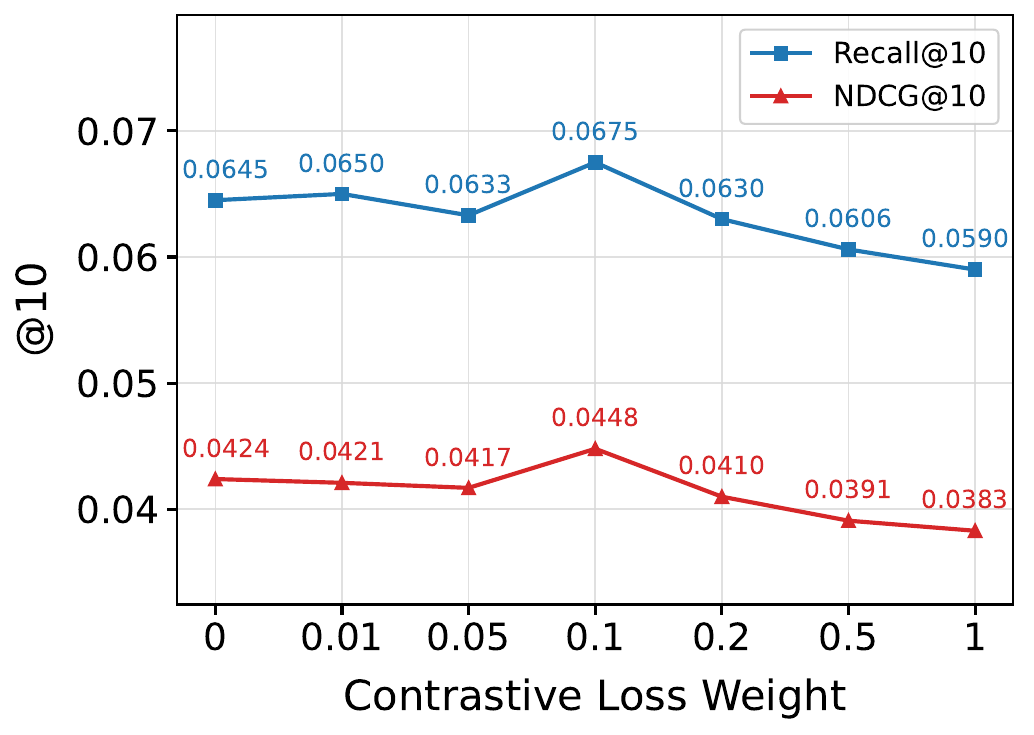}
	\quad
	\includegraphics[width=0.46\linewidth]{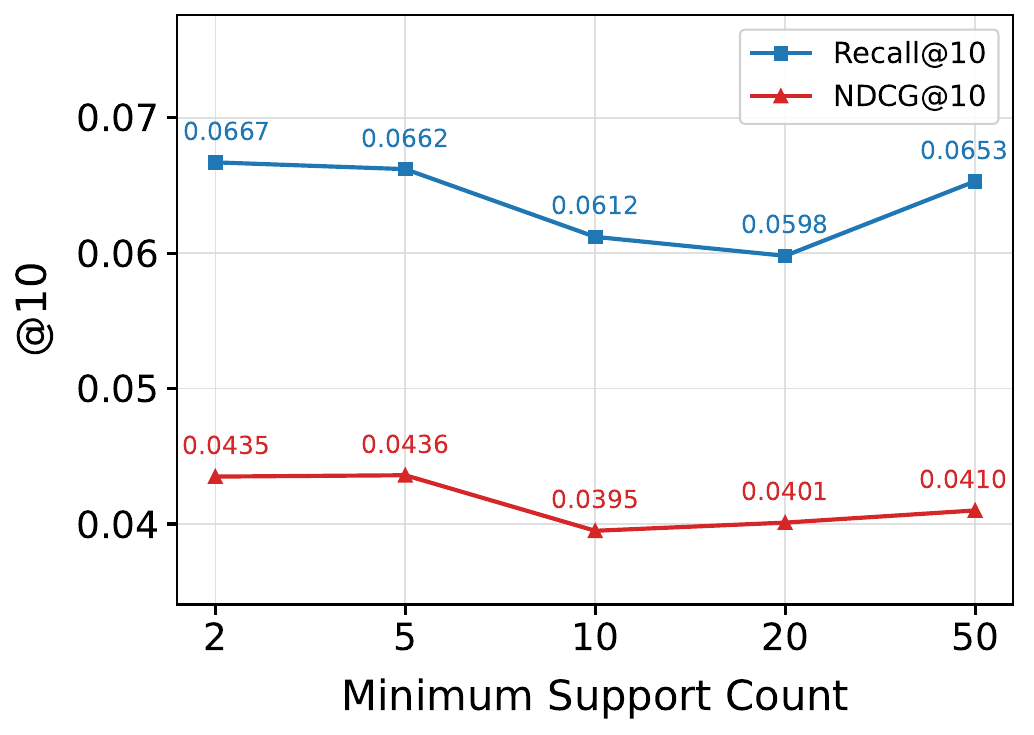}
	\caption{Parameter sensitivity analysis of {\model}.}
	\label{fig_para}
\end{figure}

Figure~\ref{fig_para} examines the impact of the contrastive loss weight $\mu$ and the minimum support count on {\model}'s performance.
1) Contrastive Loss Weight $\mu$: Performance across all datasets follows an inverted-U trend. This suggests that while moderate alignment is vital for bridging the semantic gap between experts, excessive weight over-prioritizes proximity at the expense of the model's discriminative power.
2) Minimum Support Count: The model exhibits a complex, non-monotonic response. Accuracy peaks at 2 and 5 as the graph filters spurious noise while retaining diverse signals. A decline occurs at 10 or 20 due to the knowledge loss of valuable long-tail patterns. Interestingly, performance rebounds at 50, suggesting that focusing exclusively on high-frequency core patterns provides a robust, albeit specialized, signal for recommendation.

\section{Conclusion}
\label{sec:conclusion}

In this paper, we present \textbf{TIDE}, a Time-Interval Disentangled Experts framework for Next-Basket Recommendation. We begin by empirically identifying two fundamental limitations of existing NBR methods: (1) the widespread reliance on monotonic temporal decay, which fails to capture item-specific, non-monotonic replenishment periodicities; and (2) the entanglement of habitual and exploratory user intents within a shared representation space, which suppresses the model's ability to discover novel items. To address these challenges, {\model} introduces a Hawkes-enhanced Fourier Time Encoding that jointly models personalized periodicities and dynamic intensity jumps, and a gated dual-expert architecture that explicitly decouples habitual replenishment from exploratory discovery. An InfoNCE-based contrastive alignment task further bridges the semantic gap between the two experts, ensuring a coherent fusion. Extensive experiments on four diverse real-world datasets demonstrate that {\model} consistently and significantly outperforms representative state-of-the-art NBR baselines across all metrics, validating the effectiveness of each proposed component.

Despite its strong performance, {\model} has several limitations worth noting. The global item-pattern bipartite graph remains static during inference, which may limit its ability to adapt to rapidly evolving user preferences. A natural future direction is to replace the current GNN backbone with a more scalable architecture, such as a Transformer, which offers greater flexibility for handling large-scale or dynamic graph structures. Additionally, while our contrastive alignment task effectively reduces the semantic gap between experts, the interaction between the two latent spaces could be further explored with more principled disentanglement objectives, such as information-theoretic constraints.


\begin{acks}
This work was supported in part by the National Natural Science Foundation of China under Grant 62507018; Hubei Provincial Natural Science Foundation of China under Grant 2023AFA020, JCZRMS202601468; Postdoctor Project of Hubei Province under Grant 2025HBBSHCXB056; and Fundamental Research Funds for the Central Universities under Grant CCNU25ai017.
We would like to thank the anonymous reviewers for their valuable comments.
\end{acks}


\clearpage
\bibliographystyle{ACM-Reference-Format}
\balance
\bibliography{ref}

%\appendix

\end{document}